\documentclass[11pt]{article}

\usepackage{setspace}

\usepackage{graphicx}
\usepackage{dcolumn}
\usepackage{bm}

\usepackage[T1]{fontenc}
\usepackage[latin1]{inputenc}
\usepackage{amssymb}
\usepackage{t1enc}
\usepackage{amsmath}
\usepackage{amscd}
\usepackage{mathrsfs}
\usepackage{latexsym}
\usepackage{makeidx}

\newcommand{\benumerate}{\begin{enumerate}}
\newcommand{\eenumerate}{\end{enumerate}}
\newcommand{\beq}{\begin{equation}}
\newcommand{\eeq}{\end{equation}}
\newcommand{\beqr}{\begin{eqnarray}}
\newcommand{\eeqr}{\end{eqnarray}}

\newcommand{\bitemize}{\begin{itemize}}

\newcommand{\eitemize}{\end{itemize}}

\newcommand{\der}[2]{\frac{\partial #1}{\partial #2}}

\newcommand{\dersec}[2]{\frac{\partial^{2} #1}{\partial #2^{2}}}

\begin{document}


\title{Hodographic Vortices}

\author{Antonio Moro \\
School of Mathematics,
 Loughborough University, \\
 Loughborough,
Leicestershire, LE11 3TU, UK\\
\small{email: a.moro@lboro.ac.uk}
}%

\date{}

\maketitle


\begin{abstract}
Vortices are screw phase dislocations associated with helicoidal
wave-fronts. In nonlinear optics, vortices arise as singular
solutions to the phase-intensity equations of geometric optics.
They exist for a general class of nonlinear response functions. In
this sense, vortices possess a universal character. Analysis of
geometric optics equations on the hodograph plane leads to
deformed vortex type solutions that are sensitive to the form of
the nonlinearity. The case of a Kerr type nonlinear
response is discussed as a specific example.\\
\\
PACS Numbers: 42.15.Dp, 42.65.Hw\\
Keywords: Vortices, Nonlinear Optics, Phase Singularities,
Hodograph Method
\end{abstract}


\section{Introduction}

\label{intro} Vortices are fundamental physical objects. They
appear in many different contexts: from fluid mechanics to
nonlinear optics, from superfluidity to Bose-Einstein
condensation~\cite{Walshaw,Soskin,Mazenko,Aftalion}. In 1974, J.F.
Nye and M. Berry introduced the concept of phase dislocation
obtained by interference of quasi-monochromatic wave
trains~\cite{Berry}. In optics, such phase singularities are
called optical vortices and, as suggested by Coullet {\it et
al.}~\cite{Coullet}, correspond to a state analog to vortices in
superfluidity. Vortex solitons in Kerr media have been numerically
predicted and experimentally observed by Swartzlander and Law in
1992~\cite{Swartzlander}.

The discovery of vortex solitons immediately attracted great
interest giving birth to a new branch on modern optics referred to
as nonlinear singular optics~\cite{Soskin}. Recent developments,
in this field, concern the stabilization of optical vortices by
propagation in nonlocal nonlinear media~\cite{Krolik,Segev}.

In the linear theory, phase dislocations result from an
interference phenomenon and, consequently, are not observable in
the geometric optics limit. The situation is completely different
in nonlinear regime where the geometric optics limit does not
exclude the existence of phase dislocations.

In the present Letter, we are interested in the study of possible
new singular phase solutions arising in nonlinear geometric
optics. Phase dislocations associated with such solutions have a
purely geometric nature and their properties are sensitive to the
form of the nonlinear optical response.

We emphasize that the analysis of the long wave limit is also
important for the study of the full dispersive regime. New
singular phase solutions in nonlinear geometric optics can be used
in the construction of a nontrivial ansatz for solving the
nonlinear wave equation via, for instance, variational or
numerical methods.

The Letter is organized as follows: in section 2, we derive the
phase-intensity equations for a monochromatic light beam
propagating in a generic nonlinear medium. We compute the standard
vortex (SV) associated with a helicoidal wave-front and discuss
its universal character. In section 3, a brief review of the
hodograph method is presented. In section 4, we compute, using the
hodograph method, a new family of singular phase solutions for a
Kerr type nonlinear response. We call such solutions {\em
hodographic vortices}. Hodographic vortices turn out to be a
geometric deformation of the SVs.

\section{Standard vortex}

Let us consider a stationary paraxial light beam propagating
through a weakly nonlinear medium of refractive index $n^{2}(I) =
n_{0}^{2} + \alpha^{2} n_{1}^{2} (I)$, where $I =|{\bf E}|^{2}$ is
the intensity of the electric field. We assume $n^{2}_{1}(I)$ to
be a monotonic increasing function of the intensity, vanishing at
$I=0$. The constant $n_{0}$ is the value of the refractive index
in absence of electric field. The weakly nonlinear regime is
specified by the maximal value of the intensity $I_{max}$ such
that the nonlinearity acts as a small perturbation of the
background refractive index. Such a perturbation is characterized
by the small parameter $\alpha^{2} = (n^{2}(I_{max}) -
n_{0}^{2})/n_{0}^{2}$ associated with the maximal variation of the
refractive index induced by the electric field. Slow modulations
of the electric field, propagating along the spatial direction
$z$, are described by the following nonlinear Schr\"odinger (NLS)
type equation
\begin{equation}
\label{NLS} 2i k \der{A}{Z}\; + \dersec{A}{X} + \dersec{A}{Y} +
k_{0} n_{1}^{2}(|A|^{2}) A =0,
\end{equation}
where $A=A(X,Y,Z)$ is the envelope of a linearly polarized
electric field $E = A \exp i( Z/\alpha^{2} - \omega t)$, $k_{0} =
\omega/c$ ($c \equiv $ light speed) and $k = k_{0} n_{0}$.

\noindent For our purposes, it is convenient rescaling the
 equation~(\ref{NLS}) to
the dimensionless form.

Let us $s_{0}$ be the typical spot-size, $L_{d} = k s_{0}^{2}$ the
diffraction length and $L_{nl} = 1/(k_{0} n_{1}^{2}(I_{max}))$ the
nonlinear length. Introducing the dimensionless variables $x =
X/(\sqrt{2} s_{0})$, $y = Y/(\sqrt{2} s_{0})$, $z =
Z/(2\sqrt{L_{nl} L_{d}})$, $\psi = A/\sqrt{I_{max}}$, and the
quantity $\eta = n_{1}^{2} L_{nl}/n_{0}$, the NLS
equation~(\ref{NLS}) takes the following standard form
\begin{equation}
\label{NLS_ad} i \epsilon \psi_{z} + \frac{\epsilon^{2}}{2}
\nabla^{2} \psi + \eta (|\psi|^{2}) \psi = 0,
\end{equation}
where $\epsilon = \sqrt{L_{nl}/L_{d}}$, the subscript denotes the
partial differentiation and $\nabla =
(\partial_{x},\partial_{y})$. Low dispersion/nonlinear geometric
optics limit is obtained assuming the diffraction length $L_{d}$
to be much larger than the nonlinear length $L_{nl}$, i.e.
$\epsilon \ll 1$. We perform this limit in a standard fashion,
looking for high oscillating solutions of the form $\psi = \phi
\exp (i S/\epsilon)$. Introducing the slow variables $u =
|\phi|^{2}$, $v = S_{x}$ and $w = S_{y}$ (note that by definition
$v_{y}=w_{x}$) it is straightforward to show that, in the limit
$\epsilon \to 0$, Eq.~(\ref{NLS_ad}) is equivalent to the
following dispersionless NLS type equation
\begin{subequations}
\label{quasilinear}
\begin{align}
\label{Pointyng}
&u_{z}+ \left(u v \right)_{x} + (u w)_{y} = 0 \\
\label{Eikonal1}
&v_{z} + v v_{x} + w v_{y} - \eta_{x} = 0 \\
\label{Eikonal2} &w_{z} + w w_{y} + v w_{x} - \eta_{y} = 0.
\end{align}
\end{subequations}
We refer to the monotonic function $\eta(u)$ as {\em intensity
law}. We point out that, in paraxial approximation,
Eq.~(\ref{Pointyng}) is nothing but the Poynting vector
conservation law and the equations~(\ref{Eikonal1}-\ref{Eikonal2})
are equivalent to the eikonal equation.

Let us look for stationary solutions to the system of
Eqs.~(\ref{quasilinear}) such that $S = z + F(x,y)$ and  $u =
u(x,y)$. Note that the functions $v = F_{x}$ and $w = F_{y}$ are
the transverse components of the gradient vector $(v,w,1)$ which
is orthogonal to the wavefront $(x,y,F(x,y))$. In terms of the
function $F(x,y)$, Eqs.~(\ref{quasilinear}) are equivalent to the
following equations
\begin{subequations}
\label{fluid}
\begin{align}
\label{fluid1}
&F_{x}^{2} + F_{y}^{2} = 2 \eta(u) \\
\label{fluid2} &u (F_{xx} + F_{yy}) + u_{x} F_{x} + u_{y} F_{y} =
0.
\end{align}
\end{subequations}
This system of equations is known in fluid dynamics as a model for
the two-dimensional steady, adiabatic irrotational compressible
flow. Function $F$ plays the role of the potential velocity and
$u$ the density of the fluid. Looking at $u$ as a function of
$\eta$ (this can always be done due the monotonicity of the
intensity law), and using the Eq.~(\ref{fluid1}) into the
Eq.~(\ref{fluid2}), we get the quasilinear equation of the form
\begin{equation}
\label{transverse} A F_{xx} + B F_{yy} + 2 C F_{xy} = 0,
\end{equation}
where $A = J F_{x}^{2} + 1$, $B =J F_{y}^{2} + 1$, $C = J F_{x}
F_{y}$ and $J = d(\log u(\eta))/d\eta$. The second order
equation~(\ref{transverse}) is said to be elliptic if its
discriminant $\Delta = A B - C^{2} = 4 J \eta + 1$ is strictly
positive. The ellipticity condition $\Delta > 0$ is uniformly
(i.e. for any solution) satisfied for few physically relevant
intensity laws. Important examples are the focussing Kerr-type ($u
= \eta^{\gamma}$) and logarithmic saturable ($\eta = \log (1+ u)$)
nonlinear responses. For this reason, in the following, we
restrict ourselves to the study of the elliptic case only.

It was observed in Ref.~\cite{Moro}, that, if the function $F$ is
harmonic, the corresponding wave-front is a harmonic minimal
surface. Indeed, provided $F$ to satisfy the Laplace equation
$F_{xx} + F_{yy}=0$, then $F$ solves the
equation~(\ref{transverse}) for any $J$, and consequently for any
function $u(\eta)$, if and only if it is a solution to the minimal
surfaces equation
\begin{equation}
\label{minimal} (1+F_{y}^{2}) F_{xx} + (1+ F_{x}^{2}) F_{yy} -2
F_{x} F_{y} F_{xy} = 0.
\end{equation}
Moreover, it can be proved that the only harmonic minimal surface
of the form $(x,y,F(x,y))$ is the helicoid $(x,y,\arctan
(x/y))$~\cite{Ossermann}. This singular wave-front is the SV
associated with screw type phase dislocations~\cite{Berry}.
However, in nonlinear geometric optics, SVs have a purely
geometric origin since they do not result from an interference
phenomenon. We emphasize their `universal' character due to the
fact that SVs are solutions to the system of Eqs.~(\ref{fluid})
for an arbitrary monotonic intensity law.

In the following, we construct {\em hodographic vortices} as a
family of vortex-type solutions to Eqs.~(\ref{fluid}). Unlike the
SVs, hodographic vortices are not universal in the sense specified
above. In fact, their geometric structure turns out to be
depending on the specific form of the nonlinear response.
\\
\\
\section{Hodograph Method}

The system of Eqs.~(\ref{fluid}) can be linearized by a hodograph
transformation. The hodograph method is widely used in fluid
dynamics for the study of two dimensional compressible
flows~\cite{Schreier}.

\noindent Let us introduce the {\em stream function} $\varphi$ via
equations
\begin{equation}
\label{stream} u v  = \varphi_{y}, \qquad u w = - \varphi_{x},
\end{equation}
where $v = F_{x}$ and $w = F_{y}$. Using Eqs.~(\ref{stream}) into
the identity $v_{y} = w_{x}$, one gets the following equation for
$\varphi$:
\begin{equation}
\label{stream_eq} u (\varphi_{xx}+ \varphi_{yy}) - u_{x}
\varphi_{x} - u_{y} \varphi_{y} = 0.
\end{equation}
Let us introduce polar coordinates $v = p \cos \theta$, $w = p
\sin \theta$ and suppose that the hodograph transformation $x =
x(p,\theta)$, $y =y(p,\theta)$ exists. Expanding the total
differentials $dx(p,\theta)$, $dy(p,\theta)$ and
$dF(x(p,\theta),y(p,\theta))$,
$d\varphi(x(p,\theta),y(p,\theta))$, we obtain the following set
of equations
\begin{gather}
\label{hodograph}
\begin{aligned}
x_{p} &= \frac{\cos \theta}{p} F_{p} - \frac{\sin \theta}{u p}
\varphi_{p},
 \;\; x_{\theta} = \frac{\cos \theta}{p} F_{\theta}
 - \frac{\sin \theta}{u p}
\varphi_{\theta},\\
y_{p} &= \frac{\sin \theta}{p} F_{p} + \frac{\cos \theta}{u p}
\varphi_{p},
 \; \; y_{\theta} = \frac{\sin \theta}{p} F_{\theta}
 + \frac{\cos \theta}{u p}
\varphi_{\theta}.
\end{aligned}
\end{gather}
The system of Eqs.~(\ref{hodograph}) is over-determined and its
compatibility conditions $x_{p \theta} = x_{\theta p}$ and $y_{p
\theta} = y_{\theta p}$ lead to the following equations relating
$F$ and $\varphi$
\begin{equation}
\label{phase_hod} F_{\theta} = \frac{p}{u} \varphi_{p}, \qquad
F_{p} = p \der{}{p}\left(\frac{1}{u p} \right) \varphi_{\theta}.
\end{equation}
The system of Eqs.~(\ref{phase_hod}) is also over-determined and
it is compatible if and only if $F_{p\theta} = F_{\theta p}$, i.e.
\begin{equation}
\label{stream_eq2} \der{}{p} \left(\frac{p}{u} \varphi_{p} \right)
= p \der{}{p}\left(\frac{1}{u p} \right) \varphi_{\theta \theta}.
\end{equation}
Introducing the variable $\sigma = \int_{0}^{p} u/p' \;dp'$,
Eq.~(\ref{stream_eq2}) reduces to the following {\em generalized
Tricomi} equation
\begin{equation}
\label{Tricomi_eq} \varphi_{\sigma \sigma} + K\left(\sigma \right)
\varphi_{\theta \theta} = 0,
\end{equation}
where $K\left(\sigma \right) = -p \der{}{\sigma} \left (\frac{1}{u
p} \right)$. In the derivation of Eq.~(\ref{Tricomi_eq}), it is
crucial that $u = u(\eta) = u (p^{2}/2)$ does not depend on the
variable $\theta$.

\section{Solution for a Kerr-type nonlinear response}

A focussing Kerr-type medium is specified by the intensity law of
the form $\label{Kerr_law} u = c_{0} (2\eta)^{\gamma}$, where
$c_{0}$ and $\gamma$ are certain positive constants. In this case,
we have $\sigma = c_{0} p^{2 \gamma}/(2 \gamma)$ and $u = c_{0}
p^{2 \gamma}$. Hence, the generalized Tricomi equation reduces to
the elliptic equation of the form
\begin{equation}
\label{Tricomi_Kerr} \varphi_{\sigma \sigma} +
\frac{\tilde{\gamma}}{ \sigma^{2}} \varphi_{\theta \theta} = 0,
\end{equation}
where $\tilde{\gamma} = \left(2 \gamma + 1 \right)/\left( 4
\gamma^{2} \right)$. It is {\em a priori} not obvious that the
exponent of $\sigma$ in Eq.~(\ref{Tricomi_Kerr}) does not depend
on $\gamma$.

Equation~(\ref{Tricomi_Kerr}) admits the following separable
variables solution
\begin{equation}
\label{Tricomi_sep_sol} \varphi = \left(\frac{c_{0}}{2
\gamma}\right)^{\alpha}  \; \cos \lambda \theta \; p^{2 \gamma
\alpha}
\end{equation}
where $\alpha = (1 + \sqrt{1 + 4 \tilde{\gamma} \lambda^{2}})/2$
is a positive real parameter and $\lambda$ is the separation of
variables constant and plays the role of a geometric deformation
parameter. Integration of Eqs.~(\ref{phase_hod}) gives
\begin{equation}
\label{ScaseA} F = \alpha \left(\frac{c_{0}}{2
\gamma}\right)^{\frac{\tilde{\alpha}}{2 \gamma}}  \frac{\sin
\lambda \theta}{\lambda} \; p^{\tilde{\alpha}},
\end{equation}
where $\tilde{\alpha} = 2 \gamma (\alpha - 1)$.

Assuming $\lambda^{2} \neq 1$ and integrating
Eqs.~(\ref{hodograph}), we obtain
\begin{align}
\label{mapA} x= \left (\frac{c_{0}}{2 \gamma}
\right)^{\frac{\tilde{\alpha}}{2 \gamma}} A(\theta) \; p^{\beta},
\qquad
y = \left (\frac{c_{0}}{2 \gamma} \right)^{\frac{\tilde{\alpha}}{2
\gamma}} B(\theta) \; p^{\beta},
\end{align}
where
\begin{align*}
A(\theta) &= -\frac{1}{\beta} \left[\alpha \sin \theta \cos
\lambda \theta - \frac{1+ 2 \gamma}{2 \gamma} \lambda \cos \theta
\sin \lambda \theta \right] \\
B(\theta) & = \frac{1}{\beta} \left [\alpha \cos \theta \cos
\lambda \theta + \frac{1+2 \gamma}{2 \gamma} \lambda \sin \theta
\sin \lambda \theta \right],
\end{align*}
and $\beta = 2 \gamma (\alpha -1)  - 1$.

Let us  assume that $|\lambda| < 1$. In this case, it is $\beta <
0$ and the formula~(\ref{mapA}) implies that $p \to \infty$ as
$(x,y) \to (0,0)$. Consequently, the intensity $u = c_{0}
p^{2\gamma}$ is divergent at the origin. From the
formula~(\ref{ScaseA}) it follows that the phase $S = z+ F(x,y)$
is also divergent at the origin if $\lambda \neq 0$.

We point out that the solution~(\ref{ScaseA}) is a deformation of
the SV. Indeed, we have
$$ \lim_{\lambda \to 0} F = \theta =
\arctan \left(x/y \right),
$$
On the hodograph plane, the unit vector ${\bf n} = \textup{grad}\;
S /|\textup{grad}\; S|$ normal to the wave front takes the
following remarkably simple form
\begin{equation}
\label{NormalVec} \left(\frac{p}{\sqrt{1+p^{2}}} \cos
\theta,\frac{p}{\sqrt{1+p^{2}}} \sin \theta,
\frac{1}{\sqrt{1+p^{2}}} \right).
\end{equation}
Near the origin (i.e. $p \to \infty$), we have ${\bf n} \sim (\cos
\theta, \sin \theta,0)$. Then, the unit vector normal to the wave
front is completely undetermined. Nevertheless, at the point
$(x,y) = (0,0)$ the phase~(\ref{ScaseA}) is undetermined only for
$\lambda = 0$. We call {\it hodographic vortex} a solution
possessing a point of divergent phase where the unit vector normal
to the wave-front is completely undetermined. From a physical
point of view, approaching the origin, the electric field develops
very strong oscillations even for very small $\lambda$s. Moreover,
since the intensity diverges at the origin, the $z-$axis is a
caustics. Despite the linear regime, where caustics and
dislocation lines are complementary effects~\cite{Berry}, in
nonlinear geometric optics, vortices generate caustics.
Approaching caustics, geometric optics approximation fails and the
wave corrections become important.

Note also that the phase associated with hodographic vortices
depends on the nonlinearity strength $\gamma$. This means that
hodographic vortices are sensitive to the form of the
nonlinearity.

\section{Conclusions}

Hodographic vortices arise in nonlinear geometric optics as
deformation of SVs. Unlike SVs, the refractive index distribution
induced by a hodographic vortex is sensitive to the form of the
nonlinearity.

Intriguing features of hodographic vortices stimulate the
investigation of further significant solutions to the equation
(\ref{Tricomi_Kerr}). Moreover, the study of other physically
relevant nonlinear responses will also be of interest.

We finally stress that screw type vortex solutions studied above
arise from the analysis of a particular (stationary) reduction of
the dispersionless NLS type equation~(\ref{quasilinear}). In
general, one can construct an infinite family of reductions that
are integrable via the hodograph method. Edge and edge-screw type
phase dislocations are expected to be associated with such non
stationary reductions.

\section*{Acknowledgments}

The author is grateful to B. Keyfitz and G. Ortenzi for
stimulating discussions and M. Lampis for useful comments. This
work was supported by the EPSRC grant EP/D036178/1, the FP6 Marie
Curie RTN project ENIGMA (Contract number MRTN-CT-2004-5652), and
the ESF programme MISGAM, Short Visit Grant Ref. Num. 2322.

\end{document}